\documentclass{aa}
\usepackage{graphicx,epsfig}
\usepackage{txfonts}
\usepackage{natbib}
\usepackage[colorlinks, linkcolor=blue, anchorcolor=blue, citecolor=blue]{hyperref}
\usepackage{slashbox}

\def\kms{{\rm km$\;$s$^{-1}$}}
\newcommand{\sst}[2]{\ensuremath{{#1}_{\mbox{\scriptsize{#2}}}}}

\usepackage{listings}

\begin{document}
\title{Magnetic Reconnection resulting from Flux Emergence: \
	Implications for Jet Formation in the lower solar atmosphere?}

\author{ J. Y. Ding\inst{1,2}
    \and M. S. Madjarska\inst{1}
    \and J. G. Doyle\inst{1}
    \and Q. M. Lu\inst{2}
    \and K. Vanninathan\inst{1}
    \and Z. Huang\inst{1}}

\institute{
    Armagh Observatory, College Hill, Armagh BT61 9DG, N. Ireland \\
    \email{jyd@arm.ac.uk}
    \and School of Earth and Space Sciences, University of Science and Technology 
	 of China, Hefei 230026, China }

\abstract
{}
{We aim at investigating the formation of jet-like features in the lower solar atmosphere, 
e.g. chromosphere and transition region, as a result of magnetic reconnection.}
{Magnetic reconnection as occurring at chromospheric and transition regions densities and 
triggered by magnetic flux emergence is studied using a 2.5D MHD code. The initial atmosphere is static and 
isothermal, with a temperature of $2 \times 10^4$~K. The initial magnetic field is uniform and 
vertical. Two physical environments with different magnetic field strength (25~G and 50~G) are presented. 
In each case, two sub-cases are discussed, where the environments have different initial mass 
density.}
{In the case where we have a weaker magnetic field (25~G) and higher plasma density 
($N_e=2\times~10^{11}$~cm$^{-3}$), valid for the typical quiet Sun chromosphere, a  plasma jet  would 
be observed with a temperature of 2--3~$\times~10^4$~K and a velocity as high as 40~\kms. The opposite 
case of a medium with a lower electron density ($N_e=2\times 10^{10}$~cm$^{-3}$), i.e. more 
typical for the transition region, and a stronger magnetic field of 50~G, up-flows with line-of-sight 
velocities as high as $\sim$90~\kms\ and temperatures of 6~$\times$~10$^5$~K, 
i.e. upper transition region -- low coronal temperatures, are produced. Only in the latter case, the 
low corona Fe~{\sc ix}~171~\AA\ shows a response in the jet which is comparable to the O~{\sc v} increase. }
{The results show that magnetic reconnection can be an efficient mechanism to drive plasma 
outflows in the chromosphere and transition region. The model can reproduce characteristics, 
such as temperature and velocity for a range of jet features like a fibril, a spicule, an hot 
X-ray jet or a transition region jet by changing either the magnetic field strength or the 
electron density, i.e. where in the atmosphere the reconnection occurs.}


\titlerunning{Magnetic reconnection implications for jet formation}
\maketitle

\section{Introduction}

A variety of jet-like features such as spicules, fibrils, 
surges, Ellerman bombs, EUV/X-ray jets etc., as seen in various atmospheric
regions, are observed in the solar atmosphere. Spicules 
are relatively thin, elongated jet-like structures best viewed at the solar limb in 
H$\alpha$ or Ca II images as bright features against a dark background. The 
close correlation between observed properties of limb spicules and other on-disk features 
such as mottles, fibrils and straws has prompted many authors to suggest that these may be 
counterparts of each other \citep{Tsiropoula1997}. As viewed in the optical, the classical 
spicule is observed to reach heights of 6\,500--15\,000~km \citep{Beckers1968, Withbroe1983} 
with an average lifetime of 5 minutes and average plasma velocities of 25~\kms. Recently, 
using high-resolution observations in Ca~{\sc ii}~H (3968~\AA) from the Solar Optical 
Telescope (SOT) on Hinode, \citet{depontieu07a} suggested at least two types of spicules, 
the classical spicule and a more dynamic one.

\citet{Tavabi2011} suggested four types of spicules based on their diameter, ranging 
from 0.3\arcsec\ (220 km), 0.5\arcsec\ (360 km), 0.75\arcsec\ (550 km) to 1.15\arcsec\ (850 km). 
Typically, they show a succession of upward and downward motions. The more dynamic spicule develop 
and disappear on timescales of 10--60~s, with velocities sometimes exceeding 100~\kms.  

Spicules are also seen in UV and EUV lines \citep{dere89} and thus reach at least to transition 
region temperatures. \citet{cook1984} using HRTS observations taken in C~{\sc iv}~1550~\AA\ 
showed that these structures show tilted features which was interpreted as rotational velocities 
of approximately 50 \kms. In a more recent paper, \citet{Madjarska2011} showed that this may be
better interpreted as a multi-strand structure with up-flows and down-flows.


Several authors have looked at magnetic reconnection as a driving mechanism for hot jets 
\citep[][and references therein]{jin96, innes99, galsgaard05, nishizuka08, patsourakos08, murray09, rosdahl10, pariat10}. 
\citet{roussev01c, roussev01a, roussev01b} performed 2D Magneto-hydrodynamic 
(MHD) simulations of transition region jets deriving blue-shifts of the order of 100~\kms. 
Several initial physical environments were studied. However, the plasma $\beta$ on the current 
sheet was the same in all the cases with the maximum velocity of the blue-shifted jets eventually 
reaching almost the same value in all cases, although they were different at the beginning of 
the experiments.

Various authors have suggested that spicules can be driven by waves (magnetoacoustic or Alfv\'en) 
\citep{hansteen06, depontieu07a, heggland07}, while \citet{sterling93}, \citet{karpen95}, 
and \citet{heggland09} 
have considered magnetic reconnection as a plausible candidate for driving chromospheric jets.  
In all the models mentioned above, spicules are produced but their velocities are small 
($\sim 25$~\kms). \citet{martinez10} explored 3D simulations of flux emergence, and reported on 
more dynamic features that reached coronal temperatures. In their model, large tangential discontinuities of magnetic field is necessary to 
create the conditions for accelerating the jets. These large magnetic field gradients lead to 
strong Lorentz forces, causing horizontal flows which reach a ``wall'' of strong vertical magnetic 
field. The wall squeezes the plasma and forces a large increase in the pressure, deflecting the 
horizontal flows and pushing them to move vertically.
\begin{figure*}[htbp!]
    \centering
    \resizebox{\hsize}{!}{\includegraphics[width=8cm]{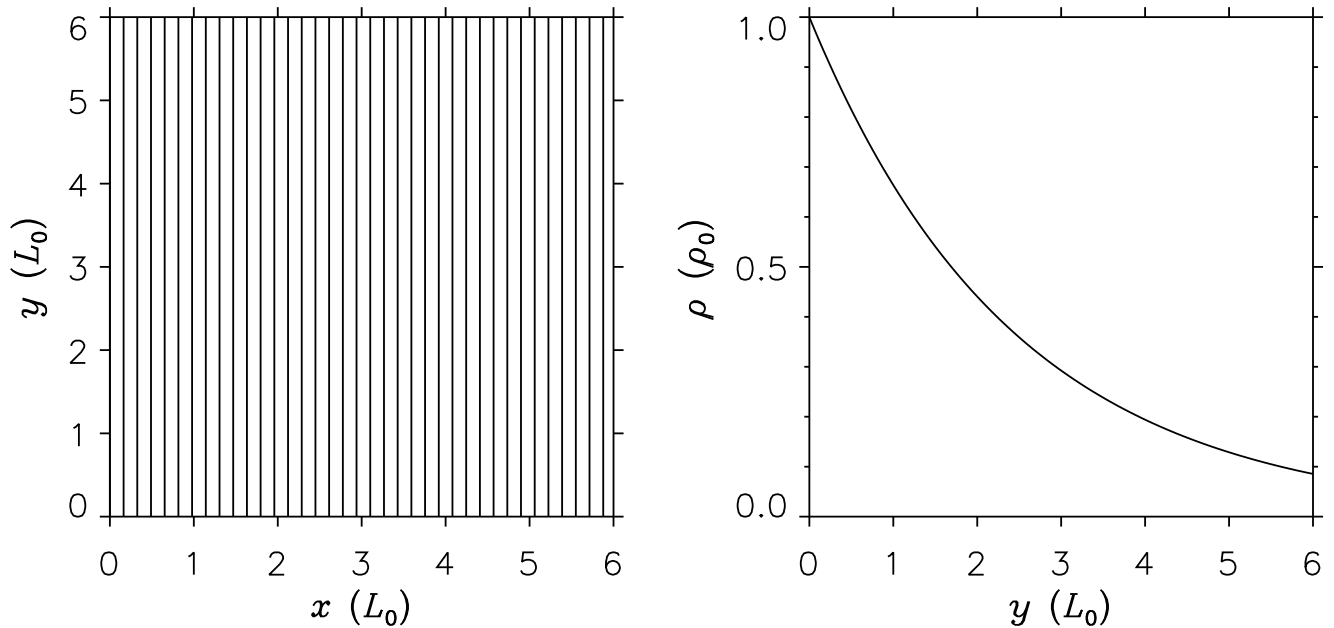}}
    \caption{Magnetic configuration of the initial potential field 
    (\emph{left}), and distribution of the initial mass density versus height
    (\emph{right}) for case A1.  }			 \label{fig1}
\end{figure*}
\begin{table}[hbp!]
    \caption{Parameters of the four initial states, namely A1, A2, B1 and B2,
    including the characteristic values of the Alfv\'en velocity ($v_A$) and 
    plasma $\beta$ in each state. 
    \label{table1}}
    \centering
    \begin{tabular}{c|cc} \hline
	\backslashbox{$\rho_{b0}$}{$\psi_{b0}$}	&   \parbox{1.5cm}{\centering{2}}	
						&   \parbox{1.5cm}{\centering{1}}	
						\\\hline
			        &   A1 	&   B1	\\
	1		    	&  $v_A=244$~\kms & $v_A=122$~\kms  \\
		&$\beta=1.1\times 10^{-2}$ & $\beta=4.4\times 10^{-2}$ \\\hline
			        &   A2 	&   B2	\\
	0.1		    	&  $v_A=772$~\kms & $v_A=386$~\kms  \\
		&$\beta=1.1\times 10^{-3}$ & $\beta=4.4\times 10^{-3}$ \\\hline
    \end{tabular}
\end{table}

\citet{pariat09} proposed a 3D model for solar polar jets, where the magnetic twist was taken 
as the jet driver. The release of magnetic twist onto open field lines by magnetic reconnection 
resulted in high-speed jets. Their work reproduced helical structures observed in some polar-hole jets 
\citep{patsourakos08}. \citet{jiang11} examined the influence of different initial
reconnection angles by using a 3D MHD model where thermal conduction and gravity were 
neglected. Fan-shaped jets moving along the magnetic guide field were obtained.

Recently, \citet{ding10} used a 2.5-dimensional resistive MHD model in Cartesian coordinates 
to investigate magnetic reconnection in the low atmosphere, e.g. chromosphere, discussing 
the implications for jet features at transition-region temperatures. They showed that faster and hotter outflows 
could be obtained if a physical environment with lower mass density was considered, however, 
the highest temperature of the plasma heated by magnetic reconnection is only $5 \times 10^5$~K 
in their study.  

Here, we expand the work of \citet{ding10} varying the electron density and field strength as 
applied to a larger range of jet features, including fibrils, spicules, chromospheric jets 
and transition region jets. This work uses a larger grid model (see later) than the previous 
study. Here, we also look at  jet formations as viewed in different parts of the lower solar atmosphere.
We derive the response of three spectral lines with formation temperatures from the lower 
transition region to the low corona, as obtained in the foot-points and in the jet itself.
 The 2.5-D resistive MHD model is briefly 
described in Section 2. Section 3 gives the numerical results and the derived line profiles.  
Conclusions and discussion are drawn in Section 4. 

\section{Physical Model and Numerical Methods}
\begin{figure*}[htbp!]
    \centering
    \resizebox{0.95\hsize}{!}{\includegraphics[width=10cm]{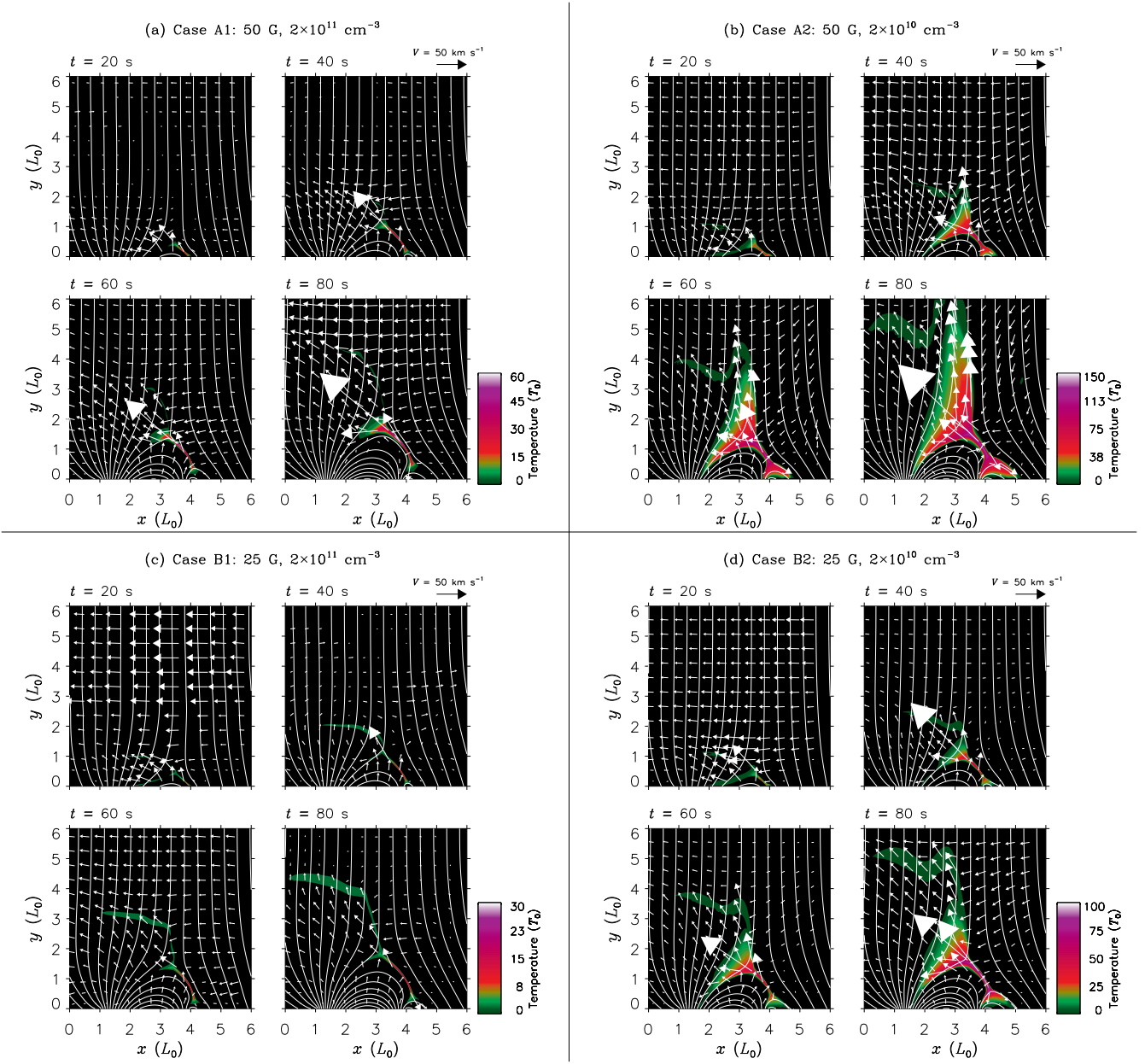}}
    \caption{Evolution of the magnetic field (\emph{solid lines}), 
    temperature (\emph{colour}), and velocity (\emph{arrows}) at four times 
	for case A1~(a), A2~(b), B1~(c), and B2~(d). 
 The characteristic parameters of magnetic field strength
	and electron number density are given above each case plot.
	Note that the colour-coding is different for each run.} \label{fig2}    
\end{figure*}
\subsection{Basic Equations}

A 2.5-dimensional resistive MHD model in Cartesian coordinates 
is used here. The MHD equations are the same as presented in \citet{ding10},
where $\rho$, $\mathbf{v}$, $\psi$, $B_z$, $T$, $p$,\ $\mathbf{j}$,\ 
$\mathbf{B}$, $\mathbf{g}$, $\eta$, $\gamma$, $\beta_0$, $Q$, $L_r$
are the mass density, flow velocity, magnetic flux function,
$z-$component of magnetic field, temperature, gas pressure, 
electric current density, magnetic field, gravitational acceleration,
dimensionless magnetic diffusivity, adiabatic index,
characteristic ratio of the gas pressure to the magnetic pressure deduced from
the basic units,
heat conduction, and radiative losses, respectively.
And, $p$, $\mathbf{j}$, $\mathbf{B}$, $Q$,
$L_r$ are explicitly expressed by
$$p=\rho T,\ \mathbf j=\bigtriangledown \times {\mathbf B},\ \textrm{and}\   
{\mathbf B} = \bigtriangledown\times \left (
	{\psi}\hat{z} \right ) + B_z \hat{z},	    \eqno (1)$$
$$Q=\bigtriangledown \cdot [T^{5/2}({\mathbf B}\cdot \bigtriangledown T){\mathbf
B}/B^2],\ \textrm{and}\ L_r = \rho^2 \Lambda (T),   \eqno (2)$$
where $\Lambda (T)$ is the radiative loss function. Here, optical depth effects are considered: 
At $T\le 2\times 10^4$~K, $\Lambda (T)$ is reduced to zero; At $2\times 10^4 < T < 10^5$~K, $\Lambda 
(T)$ suggested by \citet{mcclymont83} is used;
At $T\ge 10^5$~K, $\Lambda (T)$ calculated by \citet{cook89} is adopted.
The characteristic values taken as basic units for the mass density,
temperature, length, and magnetic field strength are: 
$\rho_0=3.34 \times 10^{-10}$ kg~m$^{-3}$,
corresponding to the electron number density $N_e=2\times 10^{11}$ cm$^{-3}$,
$T_0=10^4$~K, $L_0=500$~km, and $B_0=25$~G, respectively.
The dimensionless coefficients of the heat conduction and radiative losses are
in the same forms as given by \citet{ding10}.

The dimensionless size of the computational domain is $0\le x \le 6$ and $0 \le y \le 6$,
divided into $500\times 500$ grid points (in the \citet{ding10} work a smaller grid of 
$400\times 400$ was adopted). Uniform meshes are adopted in both $x-$ and $y-$ directions. In the present study, $x$ is the horizontal axis, and $y$ is vertical, representing the height of the solar atmosphere.
At the bottom, all quantities are fixed.  Other boundaries are treated as open,  
and all quantities are calculated in terms of equivalent extrapolation. 
A multi-step implicit scheme \citep{hu89} is used to solve the MHD equations.

\subsection{Initial State}

The initial magnetic field is a potential one taken to be in the following form
$$
	\left \{ \begin{array}{ll}
		    \psi=-\psi_{b0} x, \\
		    B_z=0,
		\end{array}
	\right . \eqno(3)
$$
where $\psi_{b0}$ is a free parameter used to control the magnetic field
strength of the initial background.
The initial magnetic field is different from that used in \citet{ding10} where
a linear force-free field with a vertical current sheet is adopted.
The initial state is assumed to be static and isothermal, with a temperature 
$T=T_i=2 \times 10^4$~K representing the atmosphere in the chromosphere.
The initial mass density is calculated by
$$\rho=\rho_{b0} \exp(-gy/2). \eqno (4)$$ 
where $\rho_{b0}$ is a free parameter used to control the mass density at
the bottom.

By choosing different values of $\psi_{b0}$ and $\rho_{b0}$, we can obtain 
four initial states represented by A1, A2, B1 and B2. The parameters
of the four initial states are listed in Table.~{\ref{table1}}.
The initial states are divided into two groups, case A ($\psi_{b0}=2$) and  
case B ($\psi_{b0}=1$), corresponding to the initial backgrounds with strong 
(50~G) and weak (25~G) magnetic field strength, respectively. Then, each
group is divided into two subgroups according to the value of $\rho_{b0}$. 
The case of $\rho_{b0}=1$ ($N_e=2\times 10^{11}$ cm$^{-3}$) represents the
initial state with high mass density, and $\rho_{b0}=0.1$ ($N_e=2\times 10^{10}$
cm$^{-3}$) with low mass density. The characteristic values of Alfv\'en
velocity ($v_A$) and the ratio of the gas pressure to magnetic pressure 
($\beta$) corresponding to each state are also listed in Table~\ref{table1}.
As the initial temperature of the background is the same for each state,
the characteristic sound speed for all the cases are the same as well,
namely 23~\kms.

Fig.~\ref{fig1} (\emph{left panel}) shows the configuration of the initial magnetic field
for the case of A1 as an example. The initial magnetic field is uniform and
vertical. Fig.~\ref{fig1} (\emph{right panel}) shows the distribution of the initial mass density 
versus height for case A1. The initial mass density is also uniform along $x$. As the radiative loss 
is reduced to zero at $T=2\times 10^4$~K, the initial state is not only in hydrostatic 
equilibrium but also in thermal equilibrium.

The magnetic configuration of case A2 is the same as case A1. The distribution
of the initial mass density for case A1 and A2 are similar, but its 
value is only one tenth of that for A2 as regards the mass density at the same
height. For cases in group B, the magnetic configurations are the same. The 
magnetic field strength in A is twice of that in B. The distribution
of the initial mass density of case B1 is the same as case A1, and that of case
B2 is the same as case A2.
\begin{figure}[htbp!]
    \centering
    \resizebox{\hsize}{!}{\includegraphics[width=10cm]{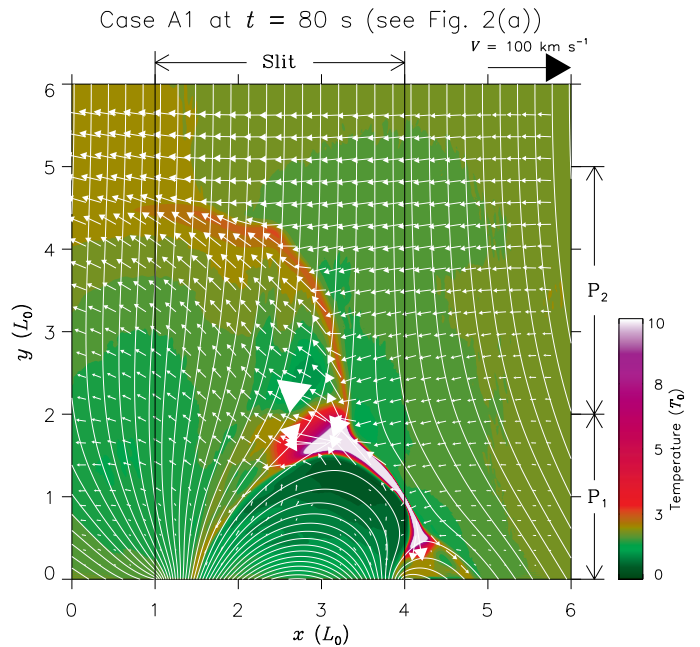}}
    \caption{Re-plot of the state at $t=80$~s in Fig.~\ref{fig2}(a). 
    The temperature is shown in a smaller color scale than in Fig.~\ref{fig2}(a).
    The two solid lines (black) denote the $x$-position of the slit, 
    and P1 and P2 denote the two $y-$position of the slit.				      \label{fig3}    }
\end{figure}
\begin{figure}[htbp!]
    \centering
    \resizebox{\hsize}{!}{\includegraphics[width=17cm]{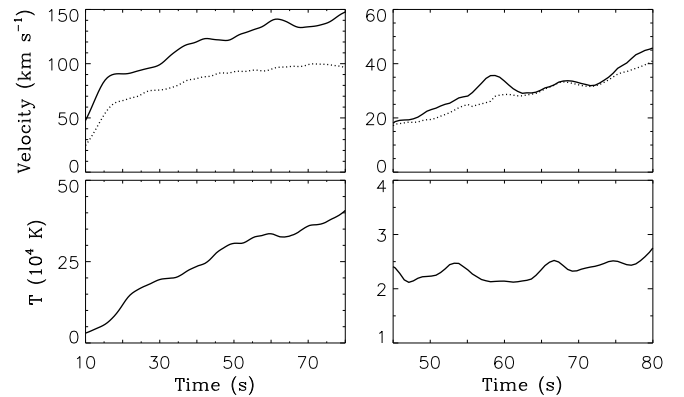}}
    \caption{Evolution of the physical quantities, e.g. maximum jet
    velocity (\emph{solid line, top row}), its line-of-sight component
    (\emph{dotted line, top row}), and plasma temperature (\emph{bottom row})
    at the location of the peak jet velocity, for case~A1.  (\emph{Left
    column}) measured at P1; (\emph{Right column}) measured at P2.  \label{fig4}
    }
\end{figure}
\subsection{New Magnetic Flux Emergence}

The emergence of new magnetic flux is implemented numerically in the same
way as presented in \citet{ding10} (see Eq.~10 therein). 
Here, the flux emergence time is
take to be 80~s as well.  The $\alpha$ parameter which controls
the magnitude and orientation of the emerging flux is taken to be $-1.2$ for
cases in group A, and $-2.4$ for cases in group B. 
The reason why $\alpha$ values are
different is to keep the ratio of the newly emerging magnetic flux to the 
total flux of the initial background the same for case A and B, so that 
reconnection between the new flux and pre-existing flux occurs at a similar height
in the two cases.
The localized resistivity introduced to 
initiate magnetic reconnection is in the same form as expressed in
\citet{ding10}.
\begin{figure*}[htbp!]
    \centering
    \resizebox{0.95\hsize}{!}{\includegraphics[width=15cm]{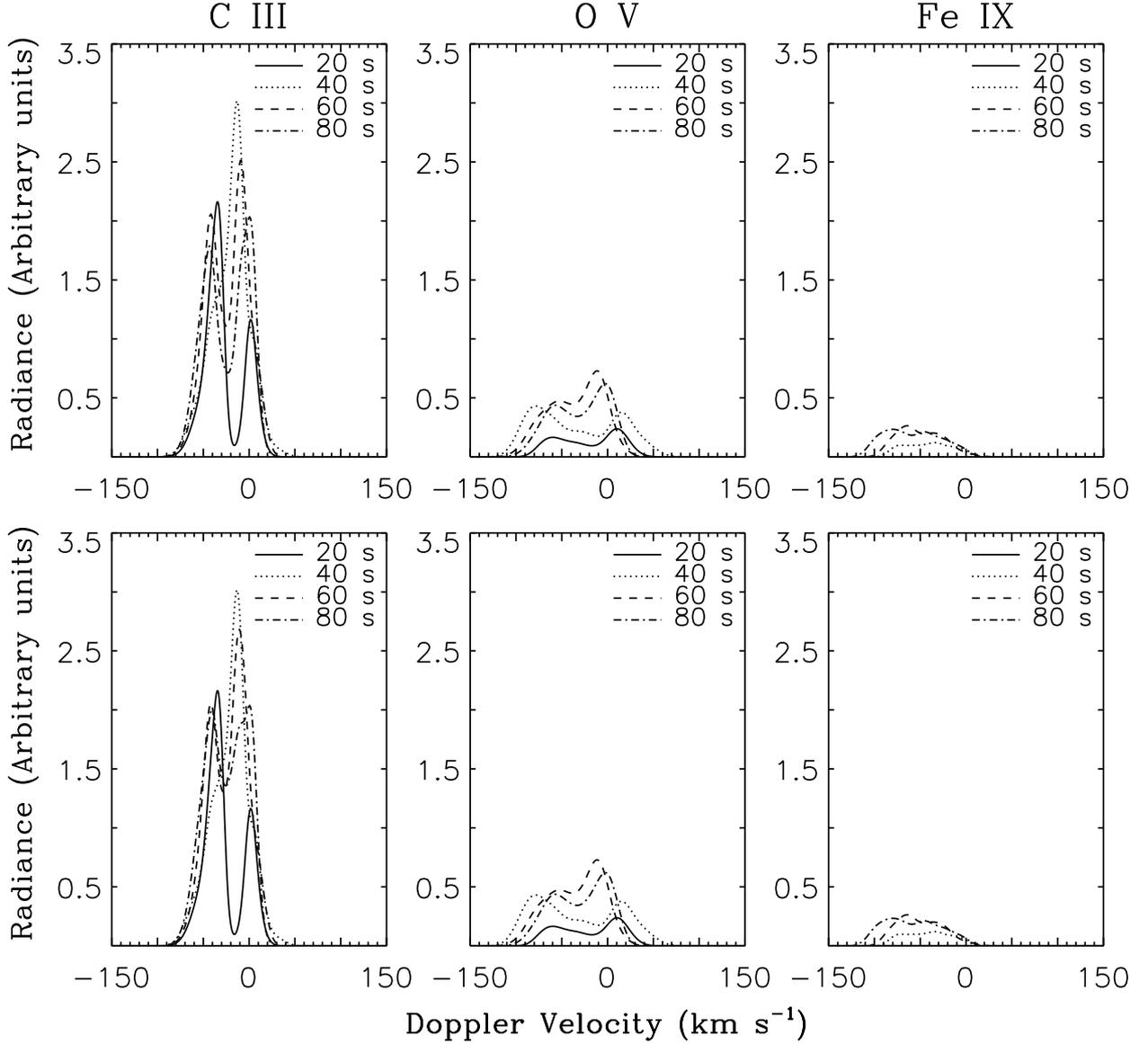}}
    \caption{Line profiles for the C~{\sc iii}~977~\AA\ line 
(\emph{left column}), the O~{\sc v}~629~\AA\ line (\emph{middle column}), and  the
Fe~{\sc ix}~171~\AA\ line (\emph{right column}) at four times for case A1.
{ Calculations are perfomed at two positions: P1 (\emph{top row}) and over
the whole region ($y=0$ -- 6, \emph{bottom row})}.
 \label{fig5} }
\end{figure*}

\subsection{Line profiles}

In order to compare the numerical simulations with observations, 
line profiles are calculated in terms of
$$
I(\lambda,t) = \int\limits_{x=1}^{x=4} \int\limits_{y=y_1}^{y=y_2} 
		\rho^2 G(T) \epsilon_T(\lambda) dx dy,\ \ \eqno(7)
$$
where $y$ represents the line-of-sight, $G(T)$ is the emission contribution function, and
$$
\epsilon_T(\lambda)\propto 
    \frac{{\exp-[\frac{(\lambda-\lambda_0-\lambda_s)}{\bigtriangleup\lambda_0}]^2}}
	{\bigtriangleup \lambda_0 \sqrt{\pi}},  \  \  \eqno(8)
	\label{eq8}
$$
is the static line profile. In the equation above, 
$ \lambda_0 $ is the rest wavelength of the resonance line,
$ \lambda_s=\frac{\lambda_0}{c} v_p$ is the Doppler shift corresponding to the 
line-of-sight velocity, $v_p$, and $\bigtriangleup \lambda_0$ is the Doppler
width of the line given by
$$
    \bigtriangleup \lambda_0=\frac{\lambda_0}{c} \sqrt{ \frac{2k_BT}{m_i} }.\ \ \eqno(9)
$$
where $c$ is the speed-of-light, $k_B$ is the Boltzmann constant, and $m_i$ is the ion mass.  
The contribution function, $G(T)$, is obtained from the ADAS database
\citep{summers09}, 
and ionization equilibrium is considered. During the calculation of line
profiles, a 1500~km wide slit, located between $1 \le x \le 4$ 
(see Fig.~\ref{fig3}), is used. When observed with spectrometers like SUMER or EIS,
line profiles from different regions of the jet can be obtained, e.g.
see Figs. 7 \& 8 in \citet{2011A&A...526A..19M}, where the small box outlined in Fig. 7 may be moved around to produce 
line profiles in a different part of the jet and/or reconnection site as seen in Fig. 8. Therefore, it is 
important that the simulation derived observables from various parts of a jet are studied and compared
with observed line profiles from the same type of regions. We, therefore, divided the simulated regions into two parts: 
P1 which covers the region between $y_1=0$ and $y_2 = 2$, i.e. over the reconnection site and its close surroundings, 
and P2 between $y_1 = 2$ and $y_2 = 5$, i.e. over the entire jet. The positions of P1 and P2 are also denoted 
in Fig.~\ref{fig3}. { Note however, that in some cases, depending on the jet temperature and density, there may be 
little or no difference in the simulated profiles at the different depths/locations, see A1 case later. Furthermore, if the 
jet was traveling directly towards the observer with no line-of-sight inclination, then deriving of the line profiles from 
different jet depths/locations would not be possible. However, jets are seen propagating in any directions on the solar disk, 
including the example outlined above, where such a comparison is possible and therefore, needs to be investigated 
both observationally and theoretically.}

In the present study, we will focus on the jets resulting from magnetic reconnection 
as seen in different parts of the solar atmosphere. Line profiles formed in the 
lower transition region (TR) are calculated for C~{\sc iii}~977~\AA~($8\times 10^4$~K), the 
upper transition region as seen in O~{\sc v}~629~\AA~($2.5\times 10^5$~K) and the lower 
corona as viewed in Fe~{\sc ix}~171~\AA~($8\times 10^5$~K).

\section{Numerical Results}

In this section, the results of magnetic reconnection triggered by newly
emerging magnetic flux are presented. Two cases are considered, case~A
and B, where the initial background magnetic field strength is different. In
each case, two sub-cases with different initial mass density are investigated. 

\subsection{Case~A1: cool chromosphere-TR jet}  \label{sec:A1}

In case~A1 we explore an environment with plasma characteristics  $\psi_{b0}=2$ and 
$\rho_{b0}=1$, i.e. B = 50~G and $N_e=2\times 10^{11}$~cm$^{-3}$, which
corresponds to high field strength typical of an active region
and high density environment.
Fig.~\ref{fig2}(a) shows the evolution of the magnetic field, temperature, 
and velocity for this case. As the new magnetic flux with negative polarity 
($\alpha<0$) emerges, the background magnetic field lines will be pushed out towards $x=0$, 
resulting in horizontal flows. A current sheet is then formed at the 
right-hand side of the newly emerging magnetic arcade. If magnetic diffusion is 
introduced into the current sheet, magnetic reconnection will occur. 
Part of the magnetic field lines expand outward (towards the upper-left in Fig.~\ref{fig2}), and the plasma 
at the right-hand side of the flux emergence region is pushed into the diffusion region.  As reconnection
goes on, the expansion of magnetic field lines propagates to higher 
regions, where plasma is dragged by the magnetic field and moves towards the left 
as seen in the figure. At the same time, magnetic field lines together with plasma from 
the right-hand boundary are driven leftwards due to pressure imbalance. This causes strong
horizontal flows at higher altitude, as seen at $t=60$~s and 80~s in
Fig.~\ref{fig2}(a). These horizontal flows will collide with the upward
outflows as a result of magnetic reconnection, and a shock is formed at the
interface. In order
to show the shock, the state at $t=80$~s in Fig.~\ref{fig2}(a) is re-plotted in
Fig.~\ref{fig3} where the temperature is plotted using a smaller colour scale. At the
interface of the horizontal flows and upward outflows, a bundle of plasma hotter
than its surrounding is clearly seen. During the magnetic reconnection, the plasma in the diffusion region is heated by
Joule dissipation. The outflows, as a result of
magnetic reconnection is bi-directional, upward and downward, along the current
concentration. In our model, only part of the down-flow  is visible because of the low X-point. 
\begin{figure}[hbp!]
    \centering
   \resizebox{\hsize}{!}{\includegraphics[width=4cm]{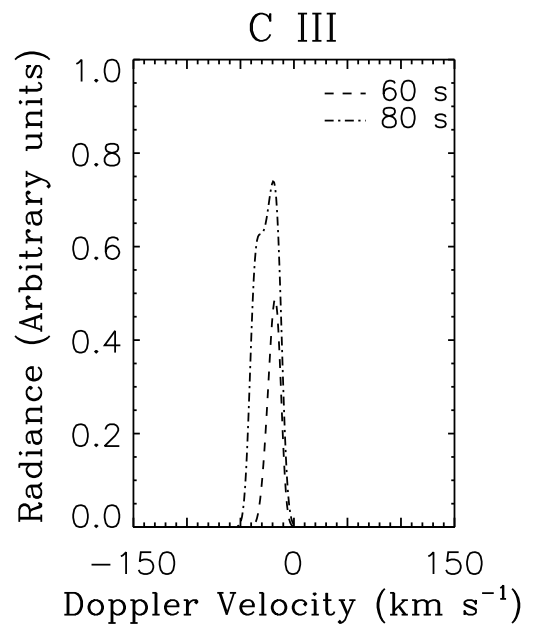}}
    \caption{Line profiles calculated at P2 for C~{\sc iii}~977~\AA\ line 
    at two times for case A1. The radiance is shown as a function of the
    Doppler shift. \label{fig6} }
\end{figure}

\begin{figure}[htp!]
    \centering
    \resizebox{\hsize}{!}{\includegraphics[width=17cm]{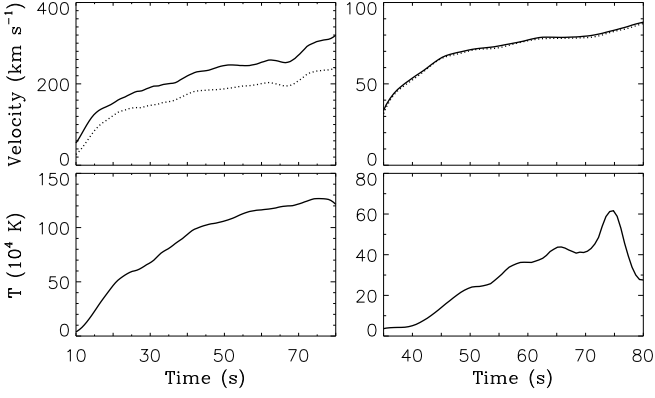}}
    \caption{Evolution of the physical quantities, e.g. maximum jet
    velocity (\emph{solid line, top row}), its line-of-sight component
    (\emph{dotted line, top row}), and the plasma temperature 
    (\emph{bottom row}) at the location of the peak jet velocity  for
    case~A2. (\emph{Left column}) measured at P1, and (\emph{right column})
    measured at P2 .	    \label{fig7}	}
\end{figure}
Fig.~\ref{fig4} (\emph{top left}) shows  the maximum jet velocity (\emph{solid
line}) and its line-of-sight component (\emph{dotted line})   
measured at P1, as a function of time. 
The corresponding  plasma temperature at the location of the maximum jet
velocity  is also shown in Fig.~\ref{fig4} (\emph{bottom left}),
as a function of time.  During the process of magnetic reconnection, the  
jet is  accelerated by magnetic tension forces, reaching a velocity of 150~\kms\
at $t=80$~s in the  P1 region, with a line-of-sight velocity of 100~\kms.
The temperature of the jets with peak velocity is about $4\times 10^5$~K
(${20\,T_i}$) at
maximum. These measurements  would correspond to a region at the 
foot-points of a chromospheric jet. The same physical quantities are also
measured at P2, where the jet velocity only reaches 45~\kms\ at maximum and the
jet temperature is $\approx\textrm{2--3}\times 10^4$~K~(${\textrm{1--1.5}\,T_i}$).  
These plasma parameters describe a short jet reaching only typical chromospheric
temperatures propagating at a speed which is comparable with observed values of
fibrils.

Fig.~\ref{fig5} (\emph{top row}) shows the C~{\sc iii}~977~\AA\ line (\emph{top left}), 
O~{\sc v}~629~\AA\ line (\emph{top middle}), and  Fe~{\sc ix}~171~\AA\ line profiles
(\emph{top right}) measured at P1 at four times, $t=20,\ 40,\ 60,\ 80$~s.
The line profiles are plotted as a function of the Doppler velocity ($v_p$), instead
of wavelength ($\lambda$), in terms of the expression $v_p=c\lambda_s/\lambda_0$.
In our line profile calculations the jets propagates towards the observer,
i.e.  the observer is at the top of the y-axis. Therefore, blue-shifted
emission corresponds to plasma moving from the bottom to the top of
the y-axis (``$+$''), while red-shifted emission is associated with plasma moving
in the opposite direction (``$-$''), i.e away from the observer. Whereas in the definition of
Doppler velocities, blue-shifts are regarded as negative values and  red-shifts as positive.
Therefore, the velocity values obtained in our simulations need
to be reversed in the calculations of line profiles, so as to get the same
expression of Doppler velocities as in the observations.
It is shown that the maximum blue-shifts reach 60~\kms\ at
$t=80$~s for the C~{\sc iii}~977~\AA\ line, and $\sim$90~\kms\ for both O~{\sc v}~629~\AA\
and Fe~{\sc ix}~171~\AA. All the line profiles show at least a two-Gaussian
structure. 

For C~{\sc iii}~977~\AA, the radiance of the red-shifted component increases during the interval 
$t=20$~s to 40~s, and then decreases. The red-shifted component centers around
velocity 0, which is mainly contributed by the plasma in and/or close to the
diffusion region. From $t=20$~s to 40~s, more plasma is heated, but to a
temperature less than $2\times 10^5$~K (which is the upper threshold of the
contribution function for C~{\sc iii}~977~\AA). As the plasma is heated to above this 
temperature, its contribution to C~{\sc iii} line becomes very weak and
negligible. The blue-shifted component comes from the plasma outside the diffusion region.
It is closer to the diffusion region at $t=20$~s, and has a strong emission.
Later, the hot region expands, but the high velocity region expands slowly 
and its mass density becomes low, leading to a decrease in the blue-shifted component at $t=40$~s.
 As more plasma is heated and accelerated, blue-shifted emission starts to increase from $t=40$~s to 60~s. 
For O~{\sc v}~629~\AA, the radiance of the highly blue-shifted components increases
from $t=20$~s to 40~s. This is because more plasma is heated to higher 
transition-region temperatures. As the temperature increases to above $2.5\times
10^5$~K (i.e. the temperature of maximum ionization ($T_{max}$) 
for O~{\sc v}~629~\AA) after $t=40$~s, the contribution of the hot plasma to the
O~{\sc v}~629~\AA\ line becomes weaker. Then, the radiance of the highly blue-shifted
components decreases, which corresponds to the increase of the radiance of the
blue-shifted components for Fe~{\sc ix}~171~\AA\ line.    

Fig.~\ref{fig6} shows the line profiles of C~{\sc iii}~977~\AA\ line measured
at P2, where the blue-shift is 20~\kms\ at $t=60$~s and reaches $\sim$30~\kms\ at
$t=80$~s. There is no C~{\sc iii} emission at $t=20$ and 40~s, because the
up-flow jets reach P2 region after $t=40$~s. In addition,
when measured at P2, the O~{\sc v}~629~\AA\ line and  Fe~{\sc ix}~171~\AA\ 
have no detectable emission. 

The results in case~A1 are summarized in Table.~\ref{table2}, where the
maximum velocity of the jet (\sst{V}{jet}), its line-of-sight component 
(\sst{V}{y}), the plasma temperature (\sst{T}{jet}) at the position of the
maximum jet velocity, and blue shifts for C~{\sc iii}~977~\AA\ (\sst{V}{C}),
O~{\sc v}~629~\AA\ (\sst{V}{O}), and Fe~{\sc ix}~171~\AA\ line (\sst{V}{Fe}) 
are calculated at two positions, P1 and P2, respectively.
All the values shown in the table are the maximum ones obtained in 80~s.
The results for other cases, namely A2, B1 and B2, which are discussed in detail 
in the following sections, are also listed similarly in Table.~\ref{table2}.

{ For the A1 case, line profiles are also calculated over the whole
height, e.g. $y=0$--6, as shown in the bottom row in Fig.~\ref{fig5}. 
There is very small difference of line profiles at P1 and that of the whole
height. In fact, the line profiles at two different depths are the same for
O~{\sc v}~629~\AA\ line and Fe~{\sc ix}~171~\AA\ lines, respectively, this is because
there is a zero contribution in the O~{\sc v} and Fe~{\sc ix} emission above the P1 region.}

\begin{table}[hbp!]
\caption{Summary of the results for all the cases, namely case A1, A2, B1 and
B2, where ``---'' means no signal. See text for details. }
\label{table2}
\centering
\begin{tabular}{c|c|cccc}
\hline
\multicolumn{2}{c|}{Quantity} &  A1 &  A2	&  B1 &  B2 \\
\hline

  & \sst{V}{jet} (\kms) &  150 & 320 & 97 & 180 \\	
  & \sst{V}{y} (\kms) &  100 & 240 & 60 & 130 \\	
\raisebox{-1.2ex}[0pt]{At P1} & \sst{T}{jet} ($T_i$) &  20 & 65 & 3.5 & 35 \\	
  & \sst{V}{C} (\kms) &  60 & 70 & 70 & 50 \\	
  & \sst{V}{O} (\kms) &  90 & 70 & --- & 50 \\	
  & \sst{V}{Fe} (\kms) &  90 & 70 & --- & 50 \\	

\hline
  & \sst{V}{jet} (\kms) &  45 & 90 & 38 & 64 \\	
  & \sst{V}{y} (\kms) &  40 & 90 & 32 & 63 \\	
  \raisebox{-1.2ex}[0pt]{At P2}  & \sst{T}{jet} ($T_i$) &  $\sim$1--1.5 & 30 & $\sim$1--1.5 & 11 \\	
  & \sst{V}{C} (\kms) &  30 & 80 & 35 & 50 \\	
  & \sst{V}{O} (\kms) &  --- & 90 & --- & 50 \\	
  & \sst{V}{Fe} (\kms) &  --- & 90 & --- & --- \\	
\hline
\end{tabular}
\end{table}

\begin{figure*}[htp!]
    \centering
    \resizebox{0.95\hsize}{!}{\includegraphics[width=17cm]{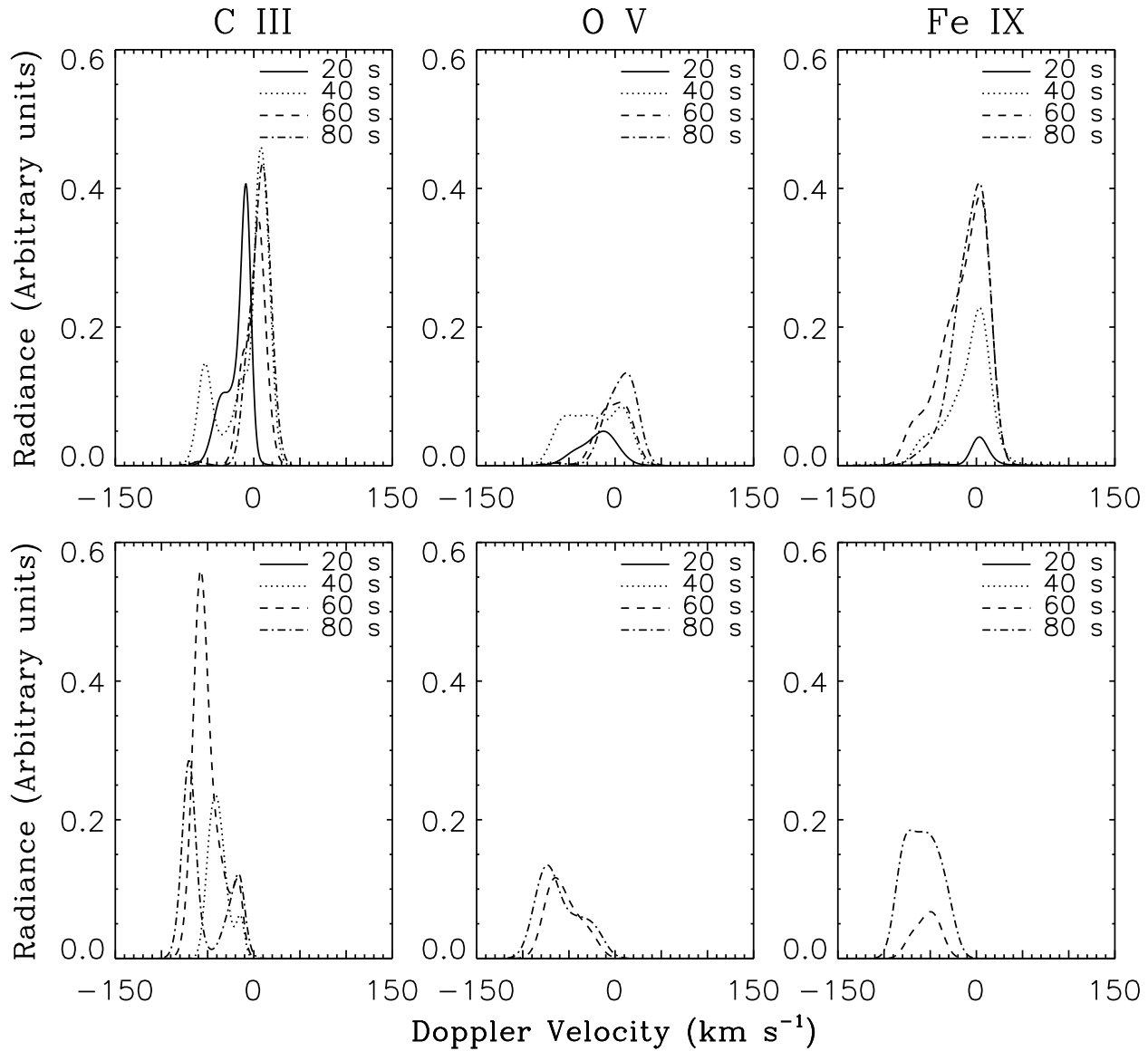}}
    \caption{Line profiles of C~{\sc iii}~977~\AA\ line (\emph{left column}),
    O~{\sc v}~629~\AA\ line (\emph{middle column}), and  Fe~{\sc ix}~171~\AA\
    line (\emph{right column}) at four times for case A2, where the line
    radiance is shown as a function of Doppler shift. Each row shows the
    observations between certain heights. \emph{Top row}: $y_1=0$ and $y_2=2$ (P1); \emph{Bottom row}:
    $y_1=2$ and $y_2=5$ (P2).				     \label{fig8} }
\end{figure*}

\subsection{Case~A2: hot transition region jet}

In the A2 case, the magnetic field background is of the same strength as in case A1, i.e. 50~G,  
but the environment has a lower electron density ($N_e=2\times 10^{10}$
cm$^{-3}$), i.e. typical of a region formed higher in the atmosphere than the A1
case. Fig.~\ref{fig2}(b) shows 
the evolution of the magnetic field, temperature, and the velocity. The plasma in the diffusion 
region is heated to higher temperatures, about $1.5 \times 10^6$~K (${75\,T_i}$) at maximum, compared with 
the case of A1. Because  an environment with lower density is considered, hotter plasma 
(i.e. transition-region temperature and above) spreads over a larger region (this corresponds 
to about 700~km wide and 3000~km high) at $t=80$~s, whereas the hotter plasma 
region is only $\sim 100$~km wide and $\sim 1000$~km high in the A1 fibril case. 
The decrease in density will increase the ratio of Joule heating to radiative losses, so 
that the plasma in the diffusion is more strongly heated. Moreover, 
the hot plasma is ejected outward, which will also heat the plasma outside the diffusion region. 
As the background atmosphere is at lower density, the energy losses of the heated hot plasma 
outside the diffusion region by radiation becomes small, so that the plasma over a larger area 
remains hot. 
\begin{figure}[htp!]
    \centering
    \resizebox{\hsize}{!}{\includegraphics[width=17cm]{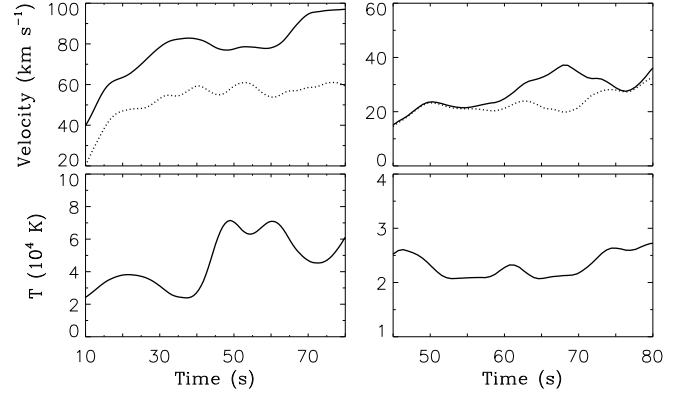}}
    \caption{Evolution of the physical quantities, e.g. maximum up-flow jet
    velocity (\emph{solid line, top row}), its line-of-sight component 
    (\emph{dotted line, top row}), and the plasma temperature
    (\emph{bottom row}) at the location of the peak jet velocity 
    for case~B1. (\emph{Left column}) 
    measured at P1; (\emph{right column}) measured at P2.	\label{fig9}	}
\end{figure}
\begin{figure}[hbp!]
    \centering
\includegraphics[scale=0.3]{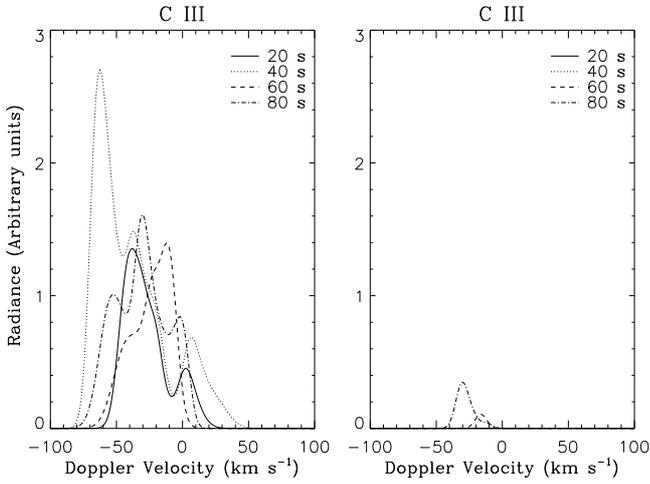}
    \caption{Line profiles of C~{\sc iii}~977~\AA\ line  observed at P1 (\emph{left})
and P2 (\emph{right}) at four times for the case of B1, where the line radiance 
    is shown as a function of Doppler shift.
    \label{fig10}	}
\end{figure}
\begin{figure}[htbp!]
    \centering
    \resizebox{\hsize}{!}{\includegraphics[width=17cm]{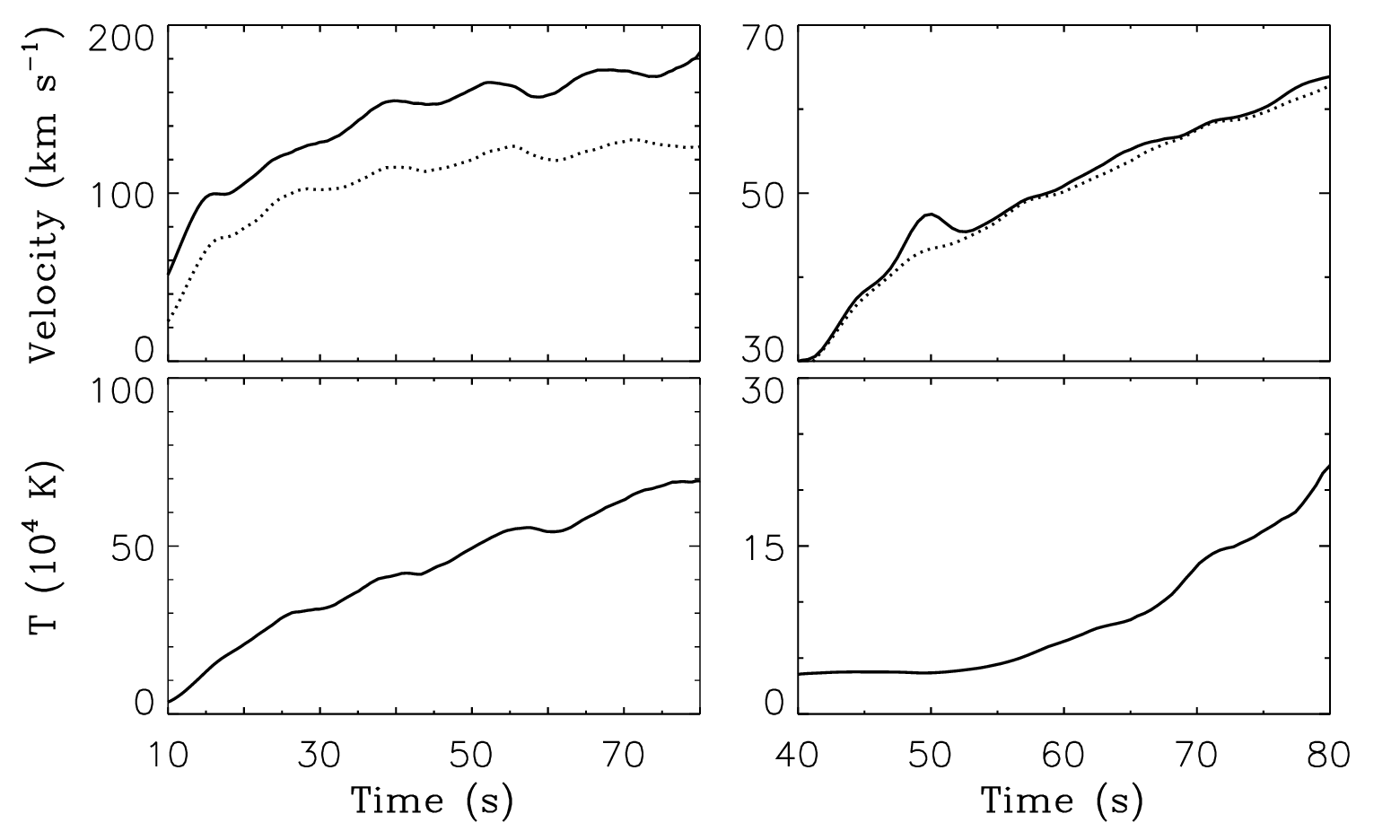}}
    \caption{Evolution of the physical quantities, e.g. maximum up-flow jet
    velocity (\emph{solid line, top row}), its line-of-sight component
    (\emph{dotted line, top row}), and the plasma temperature (\emph{bottom
    row}) at the location of the peak jet velocity for case~B2. (\emph{Left
    column}) measured at P1; (\emph{right column}) measured at P2.	\label{fig11} }
\end{figure}

Fig.~\ref{fig7} shows the maximum jet velocity (\emph{solid line, top row}), 
its line-of-sight component (\emph{dotted line, top row}),
and the plasma temperature at the peak jet velocity
(\emph{bottom row}) as a function of time, for the A2 case. 
At P1 (\emph{left column}), the up-flow jets reach 320~\kms\ at maximum,
line-of-sight component being 240~\kms. The temperatures of the up-flow jets with
maximum velocity reach about $1.3 \times 10^6$~K (${65\,T_i}$).  It seems that the outflow
jets with high velocity and high temperature are comparable to hot X-ray jets. 
When observed at P2 (\emph{right column}), however, both the velocities and temperatures are smaller:
the jet velocity and its line-of-sight component reach about 90~\kms\ at maximum, its temperature 
being about $6 \times 10^5$~K (${30\,T_i}$) at maximum. 
This means that a jet at transition region temperatures is produced.
The differences between the jet velocity and its line-of-sight component are very
small, showing that the outflows with maximum velocity are almost vertical. 
Moreover, there is a sharp drop in the temperature of the up-flow jets with peak
velocity between $t=75$ and 80~s. This is
because the outflows reach maximum velocity at higher altitudes, where
the plasma temperature is low.

Fig.~\ref{fig8} shows the line profiles of the C~{\sc iii}~977~\AA\ line
(\emph{left column}),  the O~{\sc v}~629~\AA\ line (\emph{middle column}), and
the Fe~{\sc ix}~171~\AA\ line (\emph{right column}) at four times, $t=20,\ 40,\
60,\ 80$~s, as a function of Doppler velocity. Line profiles are calculated at
two positions: P1 (\emph{top row}) and P2 (\emph{bottom row}).
Observed at P1, the maximum blue-shifts are of the order of 70~\kms\ 
for all three lines. The intensity of C~{\sc iii}~977~\AA\ and O~{\sc v}~629~\AA\
is a factor of 2--3 lower than in the A1 fibril case, while Fe~{\sc ix}~171~\AA\ is
larger. The blue-shifted components of all three lines are stronger
when observed at P2 than at P1 at $t=80$~s, whereas the red-shifted components
is absent at P2. In this case of TR temperatures and relatively strong  magnetic field, the produced 
jet is capable at reaching high TR temperatures which can peak for a short period of time at low coronal temperatures.
{ Depending on the line-of-sight, the observer may see a mixture of P1 and P2, and in instances where the jet is 
moving directly towards the observer, we will see the sum of P1 and P2 line profile.}

From an observational point of view, we should expect a small-scale brightening to occur at 
coronal temperatures but a jet will only be detected at transition region temperatures. This 
type of phenomena are already seen in multi-instrument observations and described by 
Subramanian et al. (2011, submitted). The line profiles composed of multi-component blue-shift 
and a redshift in transition region lines are similar to observable line profiles which 
describe the so-called `explosive events' (EEs). Note, however, that EE spectral line 
profiles were also associated with a surge in the plage area of active region 
\citep{2009ApJ...701..253M} as well as were linked to EUV jets (Madjarska et al. 2011, in preparation). 

\subsection{Case~B1: cool chromospheric jet}

Here, the physical environment has a weaker magnetic field strength of 25~G, and  a high 
plasma density ($N_e=2\times~10^{11}$~cm$^{-3}$). Fig.~\ref{fig2}(c)  shows the evolution 
of the magnetic field, temperature, and velocity. The physical process is the 
same as explained in Section~\ref{sec:A1}, i.e. horizontal flows and a shock interface between the inflows 
and outflows. The plasma in the diffusion region is heated to $3 \times 10^5$~K
(${15\,T_i}$).
The scale of the hot plasma region (i.e. transition-region temperature) is very small, 
50~km in width and 800~km in height, i.e. a sub-arcsec region.

In Fig.~\ref{fig9}  at P1, the velocity of the up-flows (\emph{solid line, top
left}) reaches $\sim$100~\kms\ at maximum, with a line-of-sight component 
(\emph{dotted line, top left}) of 
$\sim$60~\kms. The temperature of the up-flows with maximum velocity
is $\sim$6~$\times$~10$^4$~K (${3\,T_i}$).
At P2 (\emph{right column} in Fig.~\ref{fig9}), the maximum jet velocity 
(\emph{solid line, top right}) is only $\sim$40~\kms\ at maximum, and the jet temperature (\emph{bottom right}) is $2\textrm{--}3 \times~10^4$~K 
(${\textrm{1--1.5}\,T_i}$). Therefore, a 
cold jet that propagate with relatively low speed is produced as compared to cases A1 and A2. 

Fig.~\ref{fig10} shows the line profiles for  C~{\sc iii}~977~\AA\ at four times.
At P1 (\emph{left}) the blue-shifts reach about 60--70~\kms\ at maximum,
while at P2 (\emph{right}), the blue-shifts are only about 25--35~\kms. As the hotter plasma region is
small in scale, the emission at higher altitude is weak or absent. 
The temperatures is not high enough to 
produce sufficient O~{\sc v}~629~\AA\ and Fe~{\sc ix}~171~\AA\ line emission. A small-scale 
brightening associated with the diffusion region should be detected in transition region 
lines, while a small-scale jet could only be seen at low transition region temperatures at maximum.

\begin{figure*}[htp!]
    \centering
    \resizebox{\hsize}{!}{\includegraphics[width=17cm]{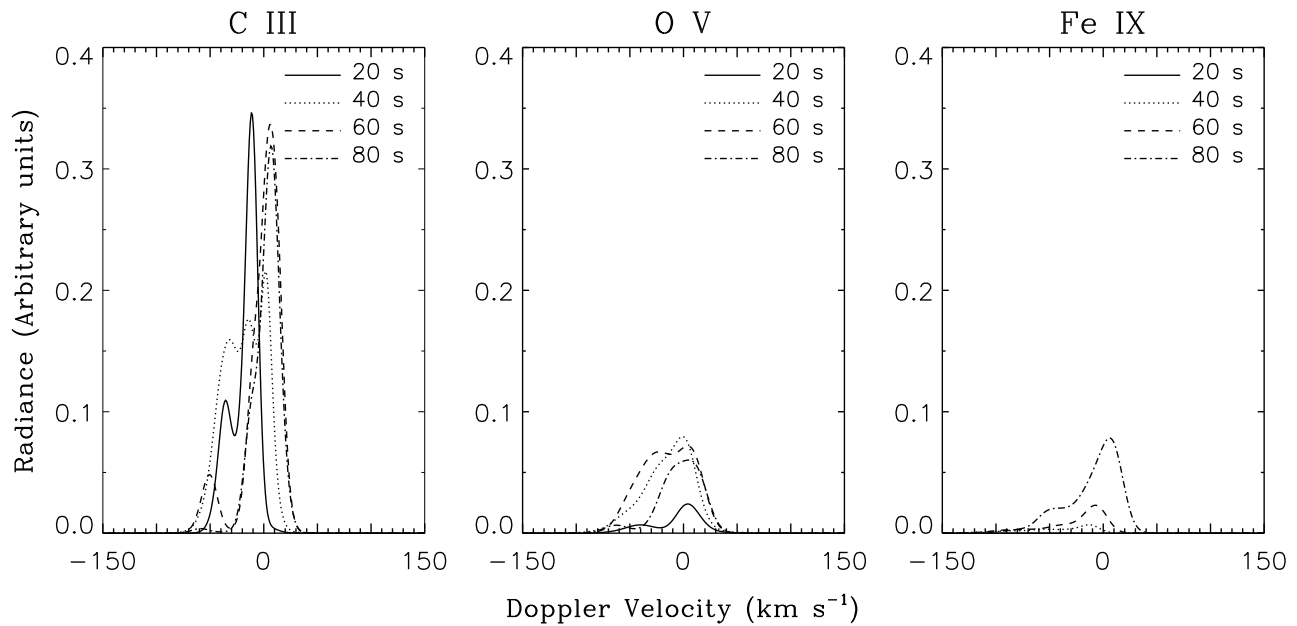}}
    \caption{Line profiles of C~{\sc iii}~977~\AA\ line (\emph{left}), O~{\sc
    v}~629~\AA\ line (\emph{middle}), and Fe~{\sc ix}~171~\AA\ line
    (\emph{right}) at four times for case B2 at P1, where the line radiance is
    shown as a function of Doppler shift. \label{fig12}	}
\end{figure*}

\subsection{Case~B2: cool chromosphere-TR jet}

\begin{figure}[hbp!]
    \centering
    \resizebox{\hsize}{!}{\includegraphics[width=17cm]{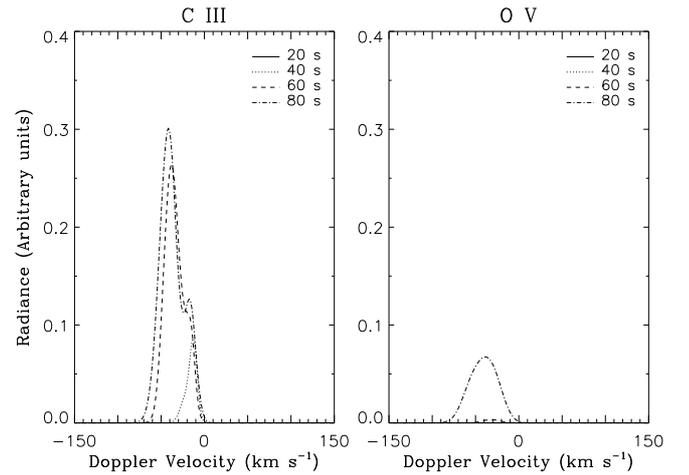}}
    \caption{Line profiles of C~{\sc iii}~977~\AA\ line (\emph{left}) and O~{\sc
    v}~629~\AA\ line (\emph{right}) at four times for case B2 at 
    P2, where the line radiance is shown as a
    function of Doppler shift. \label{fig13}	}
\end{figure}

The results of case B2 are also analyzed similarly, see Fig.~\ref{fig2}(d),
Fig.~\ref{fig11}, Fig.~\ref{fig12}, and Fig.~\ref{fig13}. In this case,
we have a lower electron density,  $N_e=2\times~10^{10}$~cm$^{-3}$, and a field
strength as in B1, i.e. 25~G. It shows that the plasma in the diffusion region
is heated up-to 1~MK (${50\,T_i}$) at maximum. Observed at P1 (\emph{left column} in
Fig.~\ref{fig11}), the maximum velocity of the
up-flows reaches 180~\kms\ at maximum, with 130~\kms\ in the line-of-sight
component. The temperature of the up-flow jets with maximum velocity reaches
$7~\times~10^5$~K (${35\,T_i}$).  
At P2 (\emph{right column} in Fig.~\ref{fig11}), both the maximum jet velocity and its line-of-sight component are
$\sim$60~\kms\ at maximum. Moreover, after $t=60$~s, the temperature of the
up-flow jets with maximum velocity increases. This is 
because the outflows get maximum velocity at low altitude, where the plasma
is heated to higher temperature. In this environment, the line profiles of all
three lines show a Doppler shift of 50~\kms, when calculated at P1
(Fig.~\ref{fig12}). At P2 (Fig.~\ref{fig13}), both C~{\sc iii}~977~\AA\ line
and O~{\sc v}~629~\AA\ line show blue-shifts of 50~\kms\ as well.
The emission in the Fe~{\sc ix}~171~\AA\ line is absent, however, 
when observed at P2. 
We should be able to register observationally an event which starts with a brightening at coronal 
temperatures, but jets will only be seen at transition region temperatures.

\section{Conclusions and Discussions}

Numerical studies on magnetic reconnection caused by new magnetic flux emergence, occurring 
at densities typical of the chromosphere and transition region, starting in an environment 
with a temperature of $2 \times 10^4$~K (${T_i}$) are presented. The magnetic field strength is set at 
25~G and 50~G, which are the typical observed values for chomospheric jets like spicules (quiet Sun) and
fibrils (active region). It is shown that the results are strongly
dependent on the physical environment: a decrease of one  order  of
magnitude of the density leads to a one time increase of the maximum jet velocity; 
{an increase of two times the magnitude of the magnetic field strength leads to the same velocity enhancement.} 
Here, the ratio of the amount of the newly emerging magnetic flux to that of the initial background is kept 
the same in all cases considered. If less new flux is emerged, or the flux emergence time is longer, 
the maximum velocity and temperature of the outflow jets will be weaker. Furthermore, both the 
inclination and non-uniformity of the initial background magnetic field will have an effect on the  
velocity and temperature of the outflow. Line profiles of two transition region lines (C~{\sc iii}~977~\AA\ and O~{\sc v}~629~\AA), 
plus the lower coronal line (Fe~{\sc ix}~171~\AA) are calculated at two positions: 
a region which covers the reconnection site and its close surroundings, and a region which covers 
the entire jet for all four cases. The simulated response in these spectral line is an important 
observable which can directly be compared with observed line profiles in various types of jets. 

In the case where we have a lower electron density ($N_e=2\times 10^{10}$~cm$^{-3}$), i.e. more 
typical for the transition region, and a stronger magnetic field of 50~G, up-flows with line-of-sight 
velocities as high as 200~\kms\ and temperatures around 1~MK (${50\,T_i}$) can be produced around the reconnection 
site. The jet at P2, however, reaches $\sim$90~\kms, but its temperatures is only 6~$\times$~10$^5$~K (${30\,T_i}$) at maximum, 
i.e. upper transition region temperatures. As a result, Fe~{\sc ix}~171~\AA\ has weak or no 
detectable emission. The opposite case of high density plasma ($N_e=2\times~10^{11}$~cm$^{-3}$), 
typical for the chromosphere and weaker magnetic field of 25~G, results in a jet with a velocity 
of $\sim$40~\kms\ and a temperature $\sim$2--3$\times$10$^4$~K 
(${\textrm{1--1.5}\,T_i}$), respectively. Therefore, a 
weak magnetic field in a high-density plasma results in a jet not higher than the typical chromospheric values.

\citet{karpen95} studied chromospheric eruptions caused by shear-induced 
magnetic reconnection. This work used a much larger box with the shear reaching the X-point
after 300~s and concluded that the dynamics of chromospheric eruptions were
dependent on the geometry, shear strength, and local resistivity. 
The present study is an important attempt to test the formation of jets originating in the lower 
solar atmosphere as produced only by magnetic reconnection triggered by new flux
emergence. Here, we do not look into the emergence rate, the emergence time or the amount of the newly emerging magnetic flux
which are important factors that could effect the results of the jets. Instead, we study
the effect of changing the field strength and electron density. The parameters have a 
significant influence on the temperature and velocity of the outflow jets.
The results demonstrate that magnetic reconnection could be an efficient mechanism to drive 
plasma outflows in the chromosphere and transition region. It is shown that if the 
magnetic field is below 50~G, the jets 
can be accelerated to relatively high velocities, but their temperatures remain in the typical chromosphere -- transition 
region range. This work suggests that intensive studies on small-scale jets and the magnetic fields 
associated with them together with a detailed spectroscopic evaluation of the jet plasma parameters 
are needed in order to confirm or reject the driving mechanism tested here and permit an evaluation of  
their contribution to the mass and energy balance in the solar atmosphere.
 
\citet{yokoy95, yokoy96} performed simulations of magnetic reconnection based on
a 2D magnetic flux emergence model. A huge amount of new magnetic flux is
emerged and evolves to the coronal heights where the reconnection with the
pre-existing background magnetic flux actually occurs. The magnetic field strength and
mass density around the diffusion region are similar to what we use in case~B1
where jets with low temperature and low velocity are obtained in our study. In the Yokoyama \& Shibata study, 
with the reconnection occurring in the corona (note that both 
heat conduction and radiative effects were neglected in their study),  
hot X-ray jets were produced. Furthermore, the emergence of new flux also
carried cold and dense plasma upward, which will then be accelerated by magnetic
tension, generating cool jets.

\citet{nishizuka08} explored MHD simulations of new flux emergence, triggering  
magnetic reconnection by adopting similar initial conditions to \citet{yokoy95,
yokoy96}, but with a hotter (1~MK) and less dense ($10^{10}$~cm$^{-3}$) corona.
Hot X-ray jets (as high as 10~MK), i.e. hotter than in \citet{yokoy95, yokoy96}, were obtained.

\citet{murray09} studied oscillatory reconnection by inserting a magnetic flux
tube below the solar surface. In their experiment the background initial magnetic field is 
$\sim$20~G, and the initial density decreases from $\sim$10$^{17}$~cm$^{-3}$ in
the lower photosphere to $\sim$10$^{9}$~cm$^{-3}$ in the upper corona.
The evolution time of the magnetic reconnection is very long
($\sim$50~mins) for the reconnection system to get close to an equilibrium state
around which reconnection reversals occur. Once the system reaches an equilibrium,
the reconnection ceases. The velocity of the jets shown in their paper are
slow ($\sim$50~\kms). However, as the reconnection occurs in the corona,
the temperature of the jets is high, reaching a few MK.
Heat conduction and radiative losses were neglected in their simulations as well.

In the present study the solar atmosphere is  fully ionized.
In the solar corona this approximation is reasonable. However, 
in the lower solar atmosphere, e.g. photosphere and chromosphere, 
the plasma is partially ionised, which means neutrals are omnipresent.
Due to the collisions with neutrals, in particular ion-neutral collisions,
a Cowling resistivity is introduced into the Ohm's law
\citep[][and references therein]{khodachenko04, arber07, soler09}.
As a result, the Joule heating is changed to 
$Q_{Joule}=\eta j_{\parallel}^{2} + \eta_{c} j_{\perp}^{2}$, where 
$\eta,\;\eta_{c},\;j_{\parallel},\;j_{\perp}$ are the Coulomb
and Cowling resistivity, and the components of the current density parallel and
perpendicular to the magnetic field, respectively. 
As far as a 2D MHD model is concerned, the electric currents are perpendicular
to the magnetic field, and thus, the Joule heating term is simplified to 
$Q_{Joule}=\eta_{c} j_{\perp}^{2} = \eta_c j^2$. In the chromosphere, $\eta_c / \eta \gg 1$ \citep{khodachenko04, leake06}, which means that the resistivity is increased
when a partially ionized plasma is considered which will result in a lower magnetic Reynolds number. \citet{litvinenko09} 
suggests the use of an enhanced value of effective resistivity, instead of the classical
resistivity ($\eta$), in studying  magnetic reconnection in the
chromosphere.

In the calculations of line profiles, ionization equilibrium
means the contribution function is mainly dependent on the electron
temperature but weakly on the electron number density. This is reasonable
for the cases where the velocity of the outflow is small, e.g. see \citet{peter04}. 
\citet{peter06} found that the assumption of ionization equilibrium is
acceptable for cases with higher velocity if the temperature gradient is
small. In our simulations, strong jets are obtained in some environments, 
e.g. case A2. Despite this, the scale of such strong jets with velocities
greater than 100~\kms\ is very small, mainly around the diffusion region. 
Over a much larger scale, the velocities of the outflow jet is small,
see top left panel (measured at P2) in Fig.~\ref{fig4}.  
Furthermore, there is a sharp density increase between the hot up-flow jet 
and the cold background plasma, which is helpful for the reduction of 
ionization and recombination times. Inside the hot up-flow region, the
temperature gradient is much flatter and in order to calculate the line emission
more precisely, it may be necessary to take into account the time-dependent
ionization in further studies, e.g. see work by \citet{sarro99, roussev01a}.

\begin{acknowledgements}

Research at Armagh Observatory is grant-aided by the N.~Ireland Dept. of 
Culture, Arts and Leisure (DCAL). This work was supported via a grant
(ST/F001843/1) from the UK Science and Technology Facilities Council.
MM and JGD thank the International Space Science Institute, Bern for
the support of the team ``Small-scale transient phenomena  and their
contribution to coronal heating''. 

\end{acknowledgements}

\bibliographystyle{aa}
\bibliography{reference}

\begin{thebibliography}{44}
\expandafter\ifx\csname natexlab\endcsname\relax\def\natexlab#1{#1}\fi

\bibitem[{{Arber} {et~al.}(2007){Arber}, {Haynes}, \& {Leake}}]{arber07}
{Arber}, T.~D., {Haynes}, M., \& {Leake}, J.~E. 2007, \apj, 666, 541

\bibitem[{{Beckers}(1968)}]{Beckers1968}
{Beckers}, J.~M. 1968, \solphys, 3, 367

\bibitem[{{Cook} {et~al.}(1984){Cook}, {Brueckner}, {Bartoe}, \&
  {Socker}}]{cook1984}
{Cook}, J.~W., {Brueckner}, G.~E., {Bartoe}, J., \& {Socker}, D.~G. 1984,
  Advances in Space Research, 4, 59

\bibitem[{{Cook} {et~al.}(1989){Cook}, {Cheng}, {Jacobs}, \&
  {Antiochos}}]{cook89}
{Cook}, J.~W., {Cheng}, C., {Jacobs}, V.~L., \& {Antiochos}, S.~K. 1989, \apj,
  338, 1176

\bibitem[{{De Pontieu} {et~al.}(2007){De Pontieu}, {Hansteen}, {Rouppe van der
  Voort}, {van Noort}, \& {Carlsson}}]{depontieu07a}
{De Pontieu}, B., {Hansteen}, V.~H., {Rouppe van der Voort}, L., {van Noort},
  M., \& {Carlsson}, M. 2007, \apj, 655, 624

\bibitem[{{Dere} {et~al.}(1989){Dere}, {Bartoe}, {Brueckner}, {Cook}, \&
  {Socker}}]{dere89}
{Dere}, K.~P., {Bartoe}, J., {Brueckner}, G.~E., {Cook}, J.~W., \& {Socker},
  D.~G. 1989, \solphys, 119, 55

\bibitem[{{Ding} {et~al.}(2010){Ding}, {Madjarska}, {Doyle}, \& {Lu}}]{ding10}
{Ding}, J.~Y., {Madjarska}, M.~S., {Doyle}, J.~G., \& {Lu}, Q.~M. 2010, \aap,
  510, A111+

\bibitem[{{Galsgaard} {et~al.}(2005){Galsgaard}, {Moreno-Insertis},
  {Archontis}, \& {Hood}}]{galsgaard05}
{Galsgaard}, K., {Moreno-Insertis}, F., {Archontis}, V., \& {Hood}, A. 2005,
  \apjl, 618, L153

\bibitem[{{Hansteen} {et~al.}(2006){Hansteen}, {De Pontieu}, {Rouppe van der
  Voort}, {van Noort}, \& {Carlsson}}]{hansteen06}
{Hansteen}, V.~H., {De Pontieu}, B., {Rouppe van der Voort}, L., {van Noort},
  M., \& {Carlsson}, M. 2006, \apjl, 647, L73

\bibitem[{{Heggland} {et~al.}(2007){Heggland}, {De Pontieu}, \&
  {Hansteen}}]{heggland07}
{Heggland}, L., {De Pontieu}, B., \& {Hansteen}, V.~H. 2007, \apj, 666, 1277

\bibitem[{{Heggland} {et~al.}(2009){Heggland}, {De Pontieu}, \&
  {Hansteen}}]{heggland09}
{Heggland}, L., {De Pontieu}, B., \& {Hansteen}, V.~H. 2009, \apj, 702, 1

\bibitem[{{Hu}(1989)}]{hu89}
{Hu}, Y.~Q. 1989, "J. Comput. Phys.", 84, 441

\bibitem[{{Innes} \& {T{\'o}th}(1999)}]{innes99}
{Innes}, D.~E. \& {T{\'o}th}, G. 1999, \solphys, 185, 127

\bibitem[{{Jiang} {et~al.}(2011){Jiang}, {Shibata}, {Isobe}, \&
  {Fang}}]{jiang11}
{Jiang}, R.~L., {Shibata}, K., {Isobe}, H., \& {Fang}, C. 2011, \apjl, 726,
  L16+

\bibitem[{{Jin} {et~al.}(1996){Jin}, {Inhester}, \& {Innes}}]{jin96}
{Jin}, S., {Inhester}, B., \& {Innes}, D. 1996, \solphys, 168, 279

\bibitem[{{Karpen} {et~al.}(1995){Karpen}, {Antiochos}, \& {Devore}}]{karpen95}
{Karpen}, J.~T., {Antiochos}, S.~K., \& {Devore}, C.~R. 1995, \apj, 450, 422

\bibitem[{{Khodachenko} {et~al.}(2004){Khodachenko}, {Arber}, {Rucker}, \&
  {Hanslmeier}}]{khodachenko04}
{Khodachenko}, M.~L., {Arber}, T.~D., {Rucker}, H.~O., \& {Hanslmeier}, A.
  2004, \aap, 422, 1073

\bibitem[{{Leake} \& {Arber}(2006)}]{leake06}
{Leake}, J.~E. \& {Arber}, T.~D. 2006, \aap, 450, 805

\bibitem[{{Litvinenko} \& {Chae}(2009)}]{litvinenko09}
{Litvinenko}, Y.~E. \& {Chae}, J. 2009, \aap, 495, 953

\bibitem[{{Madjarska}(2011)}]{2011A&A...526A..19M}
{Madjarska}, M.~S. 2011, \aap, 526, A19+

\bibitem[{{Madjarska} {et~al.}(2009){Madjarska}, {Doyle}, \& {de
  Pontieu}}]{2009ApJ...701..253M}
{Madjarska}, M.~S., {Doyle}, J.~G., \& {de Pontieu}, B. 2009, \apj, 701, 253

\bibitem[{{Madjarska} {et~al.}(2011){Madjarska}, {Vanninathan}, \&
  {Doyle}}]{Madjarska2011}
{Madjarska}, M.~S., {Vanninathan}, K., \& {Doyle}, J.~G. 2011, ArXiv e-prints

\bibitem[{{Mart{\'{\i}}nez-Sykora} {et~al.}(2010){Mart{\'{\i}}nez-Sykora},
  {Hansteen}, \& {Moreno-Insertis}}]{martinez10}
{Mart{\'{\i}}nez-Sykora}, J., {Hansteen}, V., \& {Moreno-Insertis}, F. 2010,
  ArXiv e-prints

\bibitem[{{McClymont} \& {Canfield}(1983)}]{mcclymont83}
{McClymont}, A.~N. \& {Canfield}, R.~C. 1983, \apj, 265, 497

\bibitem[{{Murray} {et~al.}(2009){Murray}, {van Driel-Gesztelyi}, \&
  {Baker}}]{murray09}
{Murray}, M.~J., {van Driel-Gesztelyi}, L., \& {Baker}, D. 2009, \aap, 494, 329

\bibitem[{{Nishizuka} {et~al.}(2008){Nishizuka}, {Shimizu}, {Nakamura},
  {Otsuji}, {Okamoto}, {Katsukawa}, \& {Shibata}}]{nishizuka08}
{Nishizuka}, N., {Shimizu}, M., {Nakamura}, T., {et~al.} 2008, \apjl, 683, L83

\bibitem[{{Pariat} {et~al.}(2009){Pariat}, {Antiochos}, \& {DeVore}}]{pariat09}
{Pariat}, E., {Antiochos}, S.~K., \& {DeVore}, C.~R. 2009, \apj, 691, 61

\bibitem[{{Pariat} {et~al.}(2010){Pariat}, {Antiochos}, \& {DeVore}}]{pariat10}
{Pariat}, E., {Antiochos}, S.~K., \& {DeVore}, C.~R. 2010, \apj, 714, 1762

\bibitem[{{Patsourakos} {et~al.}(2008){Patsourakos}, {Pariat}, {Vourlidas},
  {Antiochos}, \& {Wuelser}}]{patsourakos08}
{Patsourakos}, S., {Pariat}, E., {Vourlidas}, A., {Antiochos}, S.~K., \&
  {Wuelser}, J.~P. 2008, \apjl, 680, L73

\bibitem[{{Peter} {et~al.}(2004){Peter}, {Gudiksen}, \& {Nordlund}}]{peter04}
{Peter}, H., {Gudiksen}, B.~V., \& {Nordlund}, {\AA}. 2004, \apjl, 617, L85

\bibitem[{{Peter} {et~al.}(2006){Peter}, {Gudiksen}, \& {Nordlund}}]{peter06}
{Peter}, H., {Gudiksen}, B.~V., \& {Nordlund}, {\AA}. 2006, \apj, 638, 1086

\bibitem[{{Rosdahl} \& {Galsgaard}(2010)}]{rosdahl10}
{Rosdahl}, K.~J. \& {Galsgaard}, K. 2010, \aap, 511, A73+

\bibitem[{{Roussev} {et~al.}(2001{\natexlab{a}}){Roussev}, {Doyle},
  {Galsgaard}, \& {Erd{\'e}lyi}}]{roussev01c}
{Roussev}, I., {Doyle}, J.~G., {Galsgaard}, K., \& {Erd{\'e}lyi}, R.
  2001{\natexlab{a}}, \aap, 380, 719

\bibitem[{{Roussev} {et~al.}(2001{\natexlab{b}}){Roussev}, {Galsgaard},
  {Erd{\'e}lyi}, \& {Doyle}}]{roussev01a}
{Roussev}, I., {Galsgaard}, K., {Erd{\'e}lyi}, R., \& {Doyle}, J.~G.
  2001{\natexlab{b}}, \aap, 370, 298

\bibitem[{{Roussev} {et~al.}(2001{\natexlab{c}}){Roussev}, {Galsgaard},
  {Erd{\'e}lyi}, \& {Doyle}}]{roussev01b}
{Roussev}, I., {Galsgaard}, K., {Erd{\'e}lyi}, R., \& {Doyle}, J.~G.
  2001{\natexlab{c}}, \aap, 375, 228

\bibitem[{{Sarro} {et~al.}(1999){Sarro}, {Erd{\'e}lyi}, {Doyle}, \&
  {P{\'e}rez}}]{sarro99}
{Sarro}, L.~M., {Erd{\'e}lyi}, R., {Doyle}, J.~G., \& {P{\'e}rez}, M.~E. 1999,
  \aap, 351, 721

\bibitem[{{Soler} {et~al.}(2009){Soler}, {Oliver}, \& {Ballester}}]{soler09}
{Soler}, R., {Oliver}, R., \& {Ballester}, J.~L. 2009, \apj, 699, 1553

\bibitem[{{Sterling} {et~al.}(1993){Sterling}, {Shibata}, \&
  {Mariska}}]{sterling93}
{Sterling}, A.~C., {Shibata}, K., \& {Mariska}, J.~T. 1993, \apj, 407, 778

\bibitem[{{Summers}(2004)}]{summers09}
{Summers}, H.~P. 2004, The ADAS User Manual, version 2.6,
  http://www.adas.ac.uk/manual.php

\bibitem[{{Tavabi} {et~al.}(2011){Tavabi}, {Koutchmy}, \&
  {Ajabshirizadeh}}]{Tavabi2011}
{Tavabi}, E., {Koutchmy}, S., \& {Ajabshirizadeh}, A. 2011, \na, 16, 296

\bibitem[{{Tsiropoula} \& {Schmieder}(1997)}]{Tsiropoula1997}
{Tsiropoula}, G. \& {Schmieder}, B. 1997, \aap, 324, 1183

\bibitem[{{Withbroe}(1983)}]{Withbroe1983}
{Withbroe}, G.~L. 1983, \apj, 267, 825

\bibitem[{{Yokoyama} \& {Shibata}(1995)}]{yokoy95}
{Yokoyama}, T. \& {Shibata}, K. 1995, \nat, 375, 42

\bibitem[{{Yokoyama} \& {Shibata}(1996)}]{yokoy96}
{Yokoyama}, T. \& {Shibata}, K. 1996, \pasj, 48, 353

\end{thebibliography}

\end{document}